# Dynamic Hardness Evolution in Metals from Impact Induced Gradient Dislocation Density

Jizhe Cai, Claire Griesbach, Savannah G. Ahnen, Ramathasan Thevamaran[*]

Department of Engineering Physics, University of Wisconsin-Madison, Madison, WI 53706.
[*]Corresponding author: thevamaran@wisc.edu

**Abstract**

A clear understanding of the dynamic behavior of metals is critical for developing superior structural materials as well as for improving material processing techniques such as cold spray and shot peening. Using a high-velocity (from ∼120 m/s to 700 m/s; strain rates >$10^7$ 1/s) micro-projectile impact testing and quasistatic (strain rates: $10^{-2}$ 1/s) nanoindentation, we investigate the strain-rate-dependent mechanical behavior of single-crystal aluminum substrates with (001), (011), and (111) crystal orientations. For all three crystal orientations, the dynamic hardness initially increases with increasing impact velocity and reaches a plateau regime at hardness 5 times higher than that of at quasistatic indentations. Based on coefficient of restitution and post-mortem transmission Kikuchi diffraction analyses, we show that distinct plastic deformation mechanisms with a gradient dislocation density evolution govern the dynamic behavior. We also discover a distinct deformation regime—stable plastic regime—that emerge beyond the deeply plastic regime with unique strain rate insensitive microstructure evolution and dynamic hardness. Our work additionally demonstrates an effective approach to introduce strong spatial gradient in dislocation density in metals by high-velocity projectile impacts to enhance surface mechanical properties, as it can be employed in material processing techniques such as shot peening and surface mechanical attrition treatment.

## 1. Introduction

A fundamental understanding of the mechanical behavior and the deformation mechanisms of materials subjected to high-strain-rate dynamic loading is crucial for developing high-performance structural components in armored vehicles, spacecraft [1], and sports gears [2] that withstand impacts as well as for improving modern processing techniques such as cold spray [3], surface mechanical attrition treatment [4], and shot peening [5]. A strain-rate-dependent mechanical behavior is observed in metals when tested across quasistatic (strain rate $=10^{-3}$ 1/s) to dynamic (strain rate $=10^3$ 1/s) regimes, because of thermally activated dislocation interactions with obstacles [6]. The yield and flow stresses of metals at constant strain typically increase linearly with the logarithm of strain rate. When the strain rate is increased further beyond a threshold (strain rate $>10^3$-$10^4$ 1/s), the strain rate sensitivity of the material's mechanical properties increases dramatically, which is interpreted as the result of deformation mechanism transition to dislocation drag controlled process [7–9]. Pressure-shear plate impact tests that reach ultra-high strain rate (strain rate~ $10^5$-$10^8$ 1/s) [10,11] showed that the flow stress of pure metals increases strongly with increasing strain rate [12,13]. The microstructural origins of metallic behavior at extreme loading conditions (strain rate $>10^6$ 1/s), however, has still remained elusive because of the experimental challenges in conducting ultra-high-strain-rate dynamic tests with holistic microstructural diagnostics of the entire deformation zone on metal samples with well-defined microstructure that enables us to clearly identify structure-property relations.

Recent developments in laser-induced microprojectile impact testing (LIPIT) provide effective approach for studying the dynamic behaviors of various materials [14–18] at ultra-high strain rates with the ability to investigate the entire deformation zone and mechanisms down to atomistic resolutions. In LIPIT, a microparticle selectively launched at high-velocity (100 m/s to over 1 km/s) generates extreme loading conditions onto test materials—employed either as a projectile [14,19,20] or as target [15–17,21–24]—deforming them at far from equilibrium conditions, including ultra-high strain rates (up to ~$10^8$ 1/s) and adiabatic heating induced local temperature increase (~100s ºC) in a very short time (~a few ns). For example, face centered cubic (FCC) single-crystal silver (Ag) microcubes impacted at ~400 m/s against a rigid target has been shown to create a crystal symmetry dependent martensitic phase transformation to hexagonal close-packed (HCP) phase and a gradient nano-grained structure [14,20]. A 9R phase formation has been demonstrated in nanocrystalline aluminum (Al) thinfilms impacted by rigid microprojectiles [22]. A nanotwinning-assisted dynamic recrystallization mechanism resulting in finer nanocrystalline grain sizes was also observed in LIPIT of polycrystalline copper microparticles [24]. Hardness of polycrystalline metals (copper and iron) measured by LIPIT at ultra-high strain rate ($>10^6$ 1/s) has shown distinctly different response beyond $10^3$ to $10^4$ 1/s strain rates [15].

In contrast to polycrystalline metals, from the near perfect crystal structure without defects associated with grain boundaries, single-crystal metals exhibit remarkable high-temperature resistance to mechanical and thermal degradation [25], as well as strong physical property anisotropy [26]. This anisotropy stems from the variation of active slip systems and hardening behavior, depending on the dislocation interactions in different crystal orientations [27,28]. In dynamic regime, the anisotropic mechanical behavior of single-crystal metals with different crystal structures *i.e.* HCP [29,30], BCC [27] and FCC [28,31,32], are governed by the interplay between effects of strain rate and crystal orientations [14,20,28,30]. For example, a strong strain rate dependency of flow stress was observed in single-crystal FCC metals, *e.g.* copper, with different crystal orientations, stemming from the thermally activated motion of dislocations [31,32]. The flow stress of metals is also significantly affected by the crystal orientations with respect to the loading direction, for example, the highest flow stress is obtained at [111] direction and the lowest at [110] direction primarily governed by the characteristic slips that ensue during loading [31]. Beyond a threshold strain rate (~$10^3$ 1/s), flow stress of single-crystals exhibit increased strain rate sensitivity because of the viscous drag induced resistance on dislocation motion [33]. Dynamic hardness provides a direct quantitative measure of the resistance of metals to plastic deformation [34]. Although the quasistatic hardness and anisotropic behavior of single-crystal metals has been studied extensively using nanoindentation [35,36], the dynamic hardness, and the mechanistic origins of its strain rate sensitivity at ultra-high strain rate deformations remain elusive.

Here, we use single-crystal Al substrates as a model material to study the effects of crystal orientation and strain rate on its dynamic behavior and deformation mechanism under ultra-high strain rate impacts. Al is a lightweight FCC metal (space group $Fm\bar{3}m$; No.225) that is widely used in structural applications in its polycrystalline and nanocrystalline forms and as alloys. Using LIPIT, we selectively launched individual rigid spherical silica microparticles (diameter ~21 μm) at a wide velocity range (from ~120 to 700 m/s, strain rate ~$10^7$ 1/s) to impact the single-crystal Al substrates of three principal crystal orientations, (001), (011), and (111). We characterized the post-impact topography of impact-induced craters by optical profilometer and scanning electron microscopy (SEM) and the microstructural evolutions using transmission Kikuchi diffraction (TKD). Compared to the quasistatic hardness measured from nanoindentation, the dynamic hardness of single-crystal Al measured by LIPIT shows dependency on strain-rate and crystal orientation. The interplay between impact-induced dislocation nucleation and thermally-enhanced dislocation annihilation leads to a strain-rate-sensitive dynamic hardness evolution and a spatial gradient in dislocation density in the sample. Beyond these strain-rate-dependent plasticity regimes—elasto-plastic, fully plastic, deeply plastic regimes—we discover a distinct plastic deformation regime with strain rate insensitivity on dynamic hardness and microstructure, which we refer to as stable plastic regime.

## 2. Materials and methods

We performed the high-strain-rate impact tests in a laser-induced microprojectile impact testing system [16,17,21], shown in Fig.1(a). Single-crystal Al substrates (size: 10×10×0.5 mm and purity>99.99%, purchased from MTI Corporation) with different crystal orientations, *i.e.* (001), (011), and (111), were placed ~500 μm away from the microprojectile launch pad, which is made of a thin metal film (60 nm thick gold for low-velocity impacts ($V_i$<550 m/s) or 60 nm thick chromium for high-velocity impacts ($V_i$>550 m/s)) deposited on glass substrate followed by a 80 μm thick cross-linked polydimethylsiloxane (PDMS) layer. Monodisperse spherical silica projectiles (diameter ~21 μm, purchased from Cospheric) were drop-cast and air-dried on the launch pad and individual projectiles were selectively launched by the rapid expansion of PDMS layer from ablation of the gold layer underneath by a pulse from an Nd:YAG laser (5–8 ns pulse duration, 1064 nm). By tuning ablation laser energy, projectile velocity was varied between 120 m/s to 700 m/s. The process of projectile impacting the substrate and its rebound was imaged by a microscope camera (Allied Vision Mako G-234B) illuminated by continuous pico-second white-laser (NKT Photonics SuperK EXR-20) pulses at interval of 256.5 ns and gated by an acousto-optic modulator (ISOMET 1250C-848). The impact and rebound velocities of the projectile were calculated by dividing the distance measured between adjacent projectile snapshots in the multi-exposure image by the time interval between consecutive imaging laser pulses. Nominal strain rate of material upon impact is estimated by dividing the projectile impact velocity (120 m/s to 700 m/s) by its diameter (~21 μm, measured by optical microscope for each projectile), resulting in ~$10^6$-$10^7$ 1/s strain rates [15].

To investigate the effect of strain rate on the mechanical properties, we performed quasistatic spherical nanoindentation experiments on the single-crystal Al substrates of the three crystal orientations. Tests were performed on a Hysitron TI-950 Triboindenter equipped with a 5.78 μm radius 60º conical diamond tip. A constant nominal strain rate of $10^{-2}$ 1/s was achieved using displacement control indentation to a depth of 1 μm. Hardness values were calculated using the Oliver-Pharr method adapted for nominally spherical tip shapes [37,38].

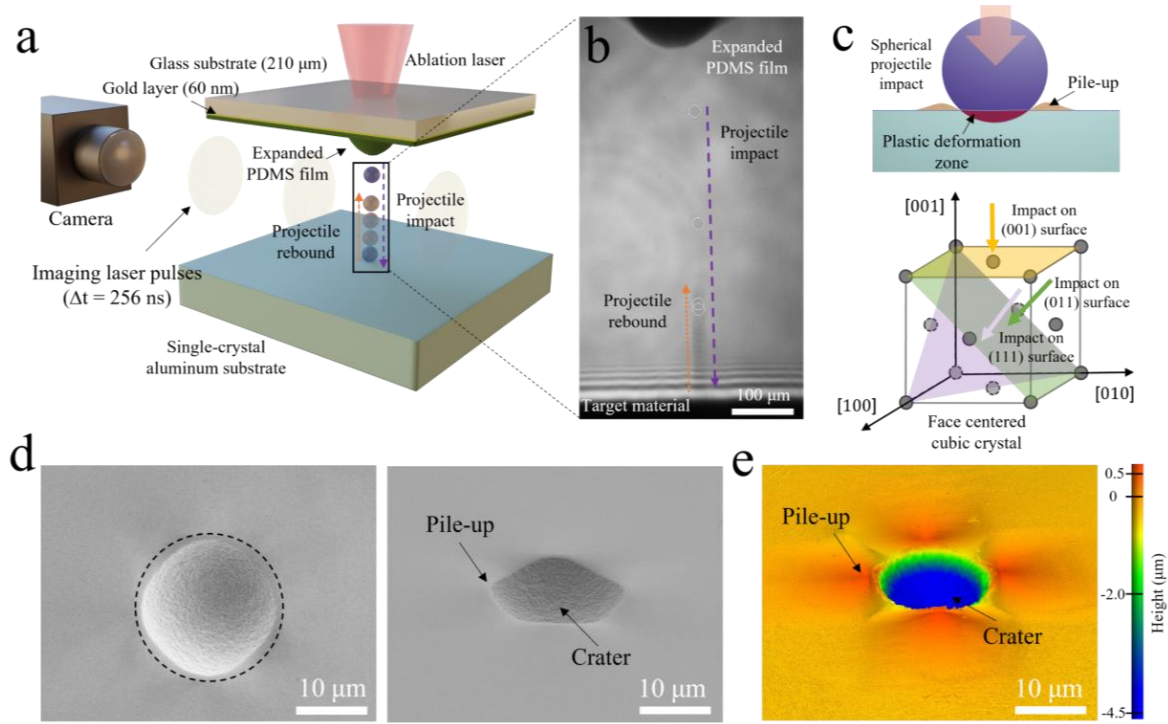

**Figure 1.** (a) Illustration of the microprojectile impact testing system for studying dynamic behaviors of single-crystal metals and (b) impact and rebound of silica projectile in a multi-exposure image. (c) Impact induced plastic deformation and surface pile-up of materials and the relationship between impact direction and crystal orientation of single-crystal Al substrates. (d) Top and tilted view SEM images of the impact induced crater by a spherical projectile (diameter: 20.8 um) impacted at 550 m/s. The dashed line indicates the size of microprojectile. (e) Surface topography of the same crater measured by an optical profilometer.

Post-impact nanostructural characterization was performed by transmission Kikuchi diffraction (TKD). Much finer spatial resolutions (<10 nm) can be achieved in this technique, which makes TKD superior to traditional EBSD methods for analyzing highly deformed materials [39–41]. Ultra-thin (~100 nm) electron-transparent lamella were formed from cross-sections extracted from beneath the impact craters using the FIB lift-out technique. Cross-sections were taken in the pile-up directions (see Fig.2). Low ion-beam dosage during the final thinning (30 kV, 50 pA) and polishing (5 kV, 100 pA) steps ensured minimal ion-beam damage and high-quality lamellae for TKD. TKD was performed using a 30 kV, 3.2 nA electron beam in an FEI Helios PFIB G4 equipped with an EDAX EBSD detector. Data analysis was performed through custom MATLAB scripts using functions from the MTEX package. The geometrically necessary dislocation density was calculated from gradients in the kernel average misorientation data using a method adapted from previous works [42,43].

## 3. Results and Discussions

### 3.1 Surface topography.

The high velocity impacts of micrometer-scale spherical projectiles normal to the single-crystal substrate surface (within $\pm 3°$) with different crystal orientations result in high-strain-rate deformation of the substrate material and the formation of impact craters, Fig.1(b, c) and Fig.S1. The topography of impact craters and quasistatic indentation imprints were examined by scanning electron microscopy (SEM) and Zygo white light optical profilometer (spatial resolution: 0.43 μm in lateral and up to 0.1 nm in depth directions) to study the effects of crystal orientation and strain rate. The plastic work done by the material under projectile impact was calculated from the kinetic energy loss of projectile during impact by measuring its impact and rebound velocities. A temperature increase in material is also expected because of the adiabatic heating from the plastic work in a very short timescale (~10 ns). For projectile impacts at velocities from 120 m/s to 700 m/s, the average final temperature ($T_{final} = T_{rise} + T_{room}$) of material in deformed region could increase between 65 °C and 670 °C (calculation in SI), with the maximum final temperature near to the melting temperature of Al (660 °C) [44]. These temperature estimates agree with the experimental observations where no melting is observed within the craters of any Al substrates we tested at impacts up to ~700 m/s (Fig.S1 in SI). The deformation response of the material, especially within the impact zone (depth = 4-2μm) (see SI), however, can very well be temperature sensitive and exhibit thermal softening at higher velocity impacts [45–47]. For instance, the flow stress of 6061-T6 aluminum alloy under dynamic loading (strain rate:$10^3$ 1/s) decreases substantially with the temperature increase, which exhibits an S-shape in flow stress-temperature curve [45] with the "knee" at ~200 °C.

As shown in Fig.1(d), in contrast to the homogeneous spherical shaped crater formed on polycrystalline metal substrates [15], the top-view SEM image of craters formed by spherical projectile impact (velocity: 550 m/s, (001) surface) shows an unusual square topography. The tilted view images of the same crater obtained by SEM and optical profilometer show a circular shaped crater with the crater rim exhibiting square shape from four pile-ups, Fig.1(d, e). The characteristic {111}<110> slip systems of the FCC Al lead to the formation of four pile-ups of the deformed material above the (001) substrate surface surrounding the impact zone. These pile-ups form the corners of the square crater rim seen in the top SEM view. Similar topography has been observed during quasistatic spherical nanoindentation of our single-crystal Al samples as well as various types of single-crystal FCC metals, including copper [48] and Ni-based alloy [36,49].

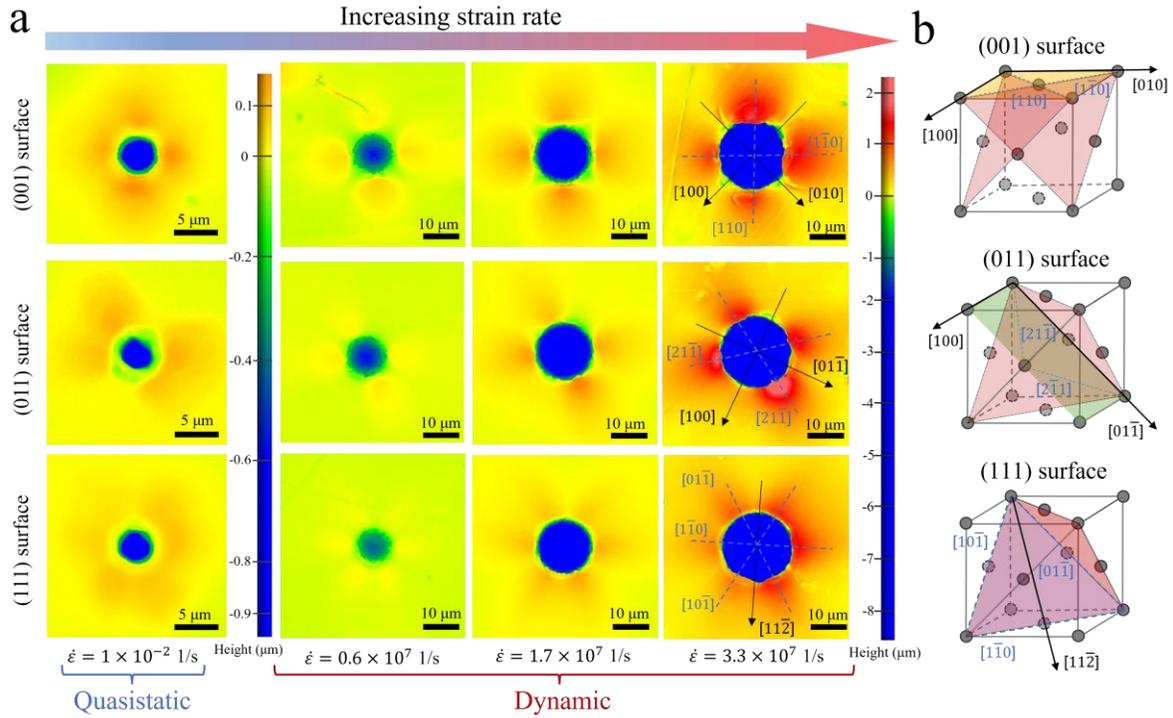

**Figure 2.** (a) Surface topography of quasistatic indentation and high-velocity projectile impacts on Al substrates with different crystal orientations. (b) Active slip planes and directions corresponding to impact along different crystal orientations of the single-crystal FCC Al substrates.

The topography of substrates with different strain rates from quasistatics ($10^{-2}$ 1/s) to dynamics ($6×10^6$ – $3.3×10^7$ 1/s, from 120 m/s to 700m/s impacts) and crystal orientations are shown in Fig.2(a). Since the indenter and the impacted projectiles have spherical geometry, the observed deviation of residual imprint on the substrate from the axisymmetry originates from the anisotropic material properties. Increasing impact velocity results in increased size and depth of the formed craters in contrast to the controlled depth in quasistatic indentation experiments. Strikingly, the deformation symmetries seen in quasistatic indentation of Al (Fig.2(a)) are preserved even in impact craters. The (001), (011), and (111) crystal orientations exhibit surface pile-ups that reflect four-fold, two-fold, and six-fold symmetry, respectively. These pile-up patterns are resultant of the anisotropic out-of-plane displacements around the impact/indentation zone because of the characteristic slips that ensue on active primary crystallographic slip systems of {111} closed packed planes slipping along <110> directions of the FCC crystal. The activation of specific slip systems leads to accumulation of material in pile-up directions elevating the crater rim above initial substrate surface [48]. As shown in Fig.2(b), the impact on (001) substrate leads to the crater topography with four-fold symmetry with pile-up along [110] and [1$\bar{1}$0] directions, the impact on (011) substrate results in two-fold symmetric topography with pile-up along [21$\bar{1}$] and [21$\bar{1}$] directions, and the impact on (111) substrate results in six-fold symmetric topography with pile-up along [01$\bar{1}$], [1$\bar{1}$0], and [10$\bar{1}$] directions. This suggests that regardless of the ultra-high strain rates up to $3.3\times10^7$ 1/s induced during impact, the slip dominant FCC single-crystal plasticity is still the governing plastic deformation mechanism.

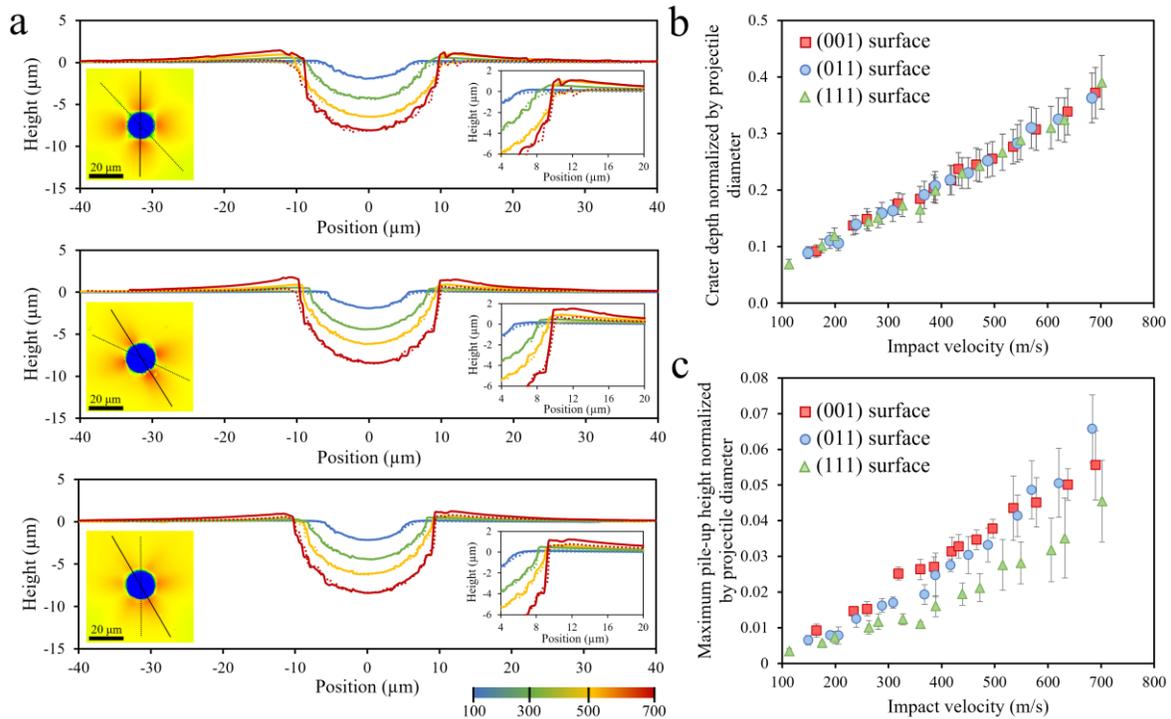

**Figure 3.** (a) Height profiles along different crystal symmetry directions (indicated by solid and dashed lines) of Al substrates of different crystal orientations following projectile impacts at different velocities (indicated by color scheme). (b) Crater depths and (c) maximum pile-up heights normalized by projectile diameters on substrates with different crystal orientations after impacts with different velocities.

The Fig.3(a) shows cross-sectional profiles of the craters on substrates with different crystal orientations impacted at different velocities. Since there is no material loss from ejection is observed during impact, based on the volume conservation in plasticity, the substrate surface is set as baseline to satisfy the conservation between the pile-up volumes and indentation volume during the surface topography measurement [15]. All crystal orientations show increase in the indentation depth and pile-up height with increasing impact velocity (Fig.3(a-c)). The surface pile-ups due to impact are formed along specific crystal orientations. A valley with much lower amount of material elevation above the substrate surface is formed between adjacent pile-ups. While the depths of craters normalized by the diameter of projectile (Fig.3(b)) do not exhibit discernible dependency on crystal orientation, a clear crystal orientation effect is observed on the maximum pile-up heights (Fig.3(c)). In contrast to the 6 pile-ups formed on (111) substrates, impacts on (001) and (011) substrates result in 4 pile-ups with larger maximum heights and the difference increases with increasing impact velocity. As the total pile-up volume—equals to the crater volume, Fig.S3—in all different substrates across entire impact velocity range do not show much difference, this maximum height difference results from the smaller number of impact-induced pile-ups on (001) and (011) substrates compared to (111) substrate.

### 3.2 Projectile rebound behavior.

To characterize the dynamic deformation regimes of single-crystal Al, we employ the semi-empirical plasticity model developed by Wu, et al. [50,51], which assumes an inverse power-law between the coefficient of restitution (COR)—the rebound velocity normalized by the impact velocity—and the impact velocity, $COR \propto V_i^{-m}$. The power-law exponent depends on three distinct sub-regimes of plasticity at which the rebounding initiates—elasto-plastic, fully plastic, and deeply plastic—that emerges as a function of projectile penetration depth, *i.e.* crater depth, into the substrate [52,53]. Initially, the scaling law follows $COR \propto V_i^{-1/4}$ in the "elasto-plastic" regime, typically in the order of 10 m/s impact velocity [53]. As the plastic deformation reaches the "fully plastic" deformation, the scaling law follows $COR \propto V_i^{-1/2}$ and further increase in impact velocity leads in to the "deeply plastic" rebound behavior, which follows a different scaling law of $COR \propto V_i^{-1}$ [53]. The relationships between $COR$ and impact velocity of single-crystal Al with different crystal orientations are plotted in a log-log plot in Fig.4. In all three crystal orientations, the rebound responses corresponding to impact velocities from 200 m/s up to 700 m/s follow a scaling law with the power-law exponent of -1. This suggests that the rebound initiation of projectiles in single-crystal Al substrate entered the "deeply plastic" regime despite the crystal orientation. Similar behavior has been observed in tungsten projectiles impacted onto copper substrates as well [52]. The $COR$ of impact onto substates of all crystal orientations at velocity lower than 200 m/s, however, deviates from the deeply plastic rebound scaling law and instead follows the "fully plastic" rebound scaling of $COR \propto V_i^{-1/2}$. This suggests that the transition from fully plastic rebound initiation to deeply plastic rebound initiation in single-crystal Al occurs at an impact velocity of ~200 m/s for all crystal orientations. Previous models have hypothesized that this transition, which depends on the impact velocity, arises from the change in rebound initiation location from the projectile/substrate interface to the inner substrate underneath the interface [52,53]. This process is directly related to the achievement of maximum stored elastic strain energy in substrate, governed by the velocity reversal inside the substrate, and the following energy release induced rebound of the projectile [53].

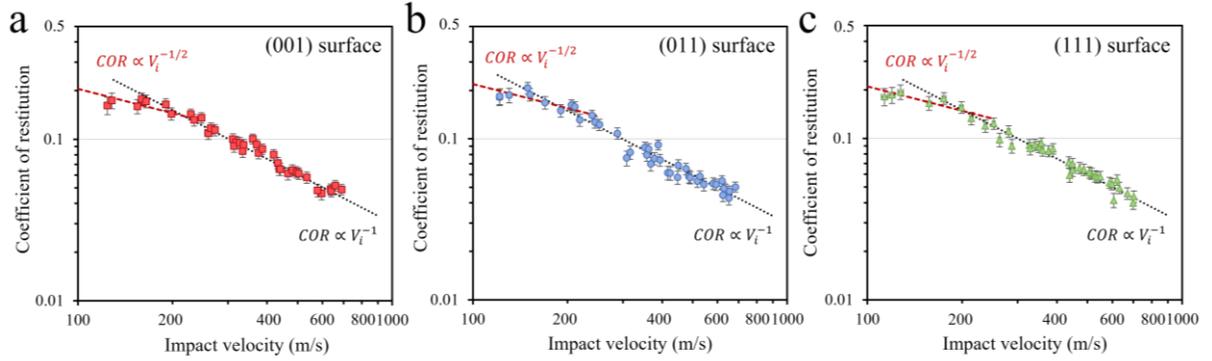

**Figure 4.** Coefficient of restitution (COR) of projectile impacts on single crystal substrates, (001), (011), and (111) surfaces with different projectile velocities shown in log-log plots. The dashed lines show curve fit to the scaling law of $COR \propto V_i^{-m}$ with $m = 1$ (black) and $m = 1/2$ (red).

### 3.3 Microstructural evolution under impact

Our TKD nanostructure analysis provides further insights into the strain rate dependent deformation mechanisms of single-crystal Al under impact. We selected impact craters formed from projectiles with low ($V_i$ = 125 m/s), medium ($V_i$ = 353 m/s), and high ($V_i$ = 550 m/s) velocity on the (001) Al substrate for these nanostructural characterizations. Clear conical influence zones can be identified by the visible boundaries marking the end in the pronounced crystal rotation (Fig.5(a–c)) and dislocation density distribution (Fig.5(d-f)). The most crystal rotation occurs in the upper corners near the boundaries of the impact crater to accommodate the formation of pile-ups, especially at higher velocity case with higher pile-up (Fig.5(c), Fig.3(c)). Gradual changes in the crystal orientation are evident, which becomes more extreme with higher impact velocity. This is also apparent in the pole figures shown in Fig.5(g–i): although the same prominent orientations emerge for all velocities, the texture becomes more diffuse with increasing impact velocity. After projectile impacts (Fig.5(a–c)), no distinct grains can be identified, even in the highest impact velocity case (550 m/s, Fig.5(c)). Although impact-induced grain refinement is commonly observed in many materials, it is not surprising that we do not see grain refinement here due to the high stacking fault energy of Al and the lack of unconstraint plastic flow that can occur when an Al microparticle is impacted against a rigid target [19]. Accompanied by the crystal orientation change, impact-induced dislocation nucleation occurs near the surface to accommodate the crystal rotation forming pile-ups, which elevates the dislocation density above the original bulk density. For the lowest impact velocity (125 m/s), the dislocation density is only slightly higher at the surface than in the bulk, revealing that minimal impact-induced dislocation nucleation has occurred. The minimal plastic deformation that occurs at low impact velocities may be accommodated through activation of existing dislocations and sources. This elevation in dislocation density is much higher for the medium and high impact velocity impacts compared to low velocity impact (Fig.5(j)). A transition in the dominant dislocation generation mechanism from Frank-Read sources to homogeneous nucleation is reported to occur at a strain rate of ~$5\times10^7$ s$^{-1}$ in Al [54], which may explain the dramatic increase in dislocation density near the surface with higher velocity impacts. This also marks the transition from the so called "fully-plastic" to "deeply-plastic" zones in the *COR* dependency on impact velocity, as the large plastic deformation forming pile-ups significantly affects the rebound mechanics. Dislocation nucleation near the surface continues to elevate the dislocation density with higher impact velocities until a critical impact velocity at which the dislocation density saturates (< 353 m/s). Such dislocation density saturation is likely due to dislocation annihilation mechanisms activated through local temperature rises. At high velocity impact, the local temperature (~500 °C) in the affected zone (depth ~2.4 μm) within substrate is higher than the temperature from medium velocity impacts (~270 °C), which results in the lower dislocation density from surface to ~4 μm depth region, due to the greater contribution from thermally-enhanced dislocation migration and annihilation in this area [54]. With further depth, the thermal effect from local temperature rise becomes less important than the dislocation density nucleation increase from impact, leading to the higher dislocation density in material from high impact velocity.

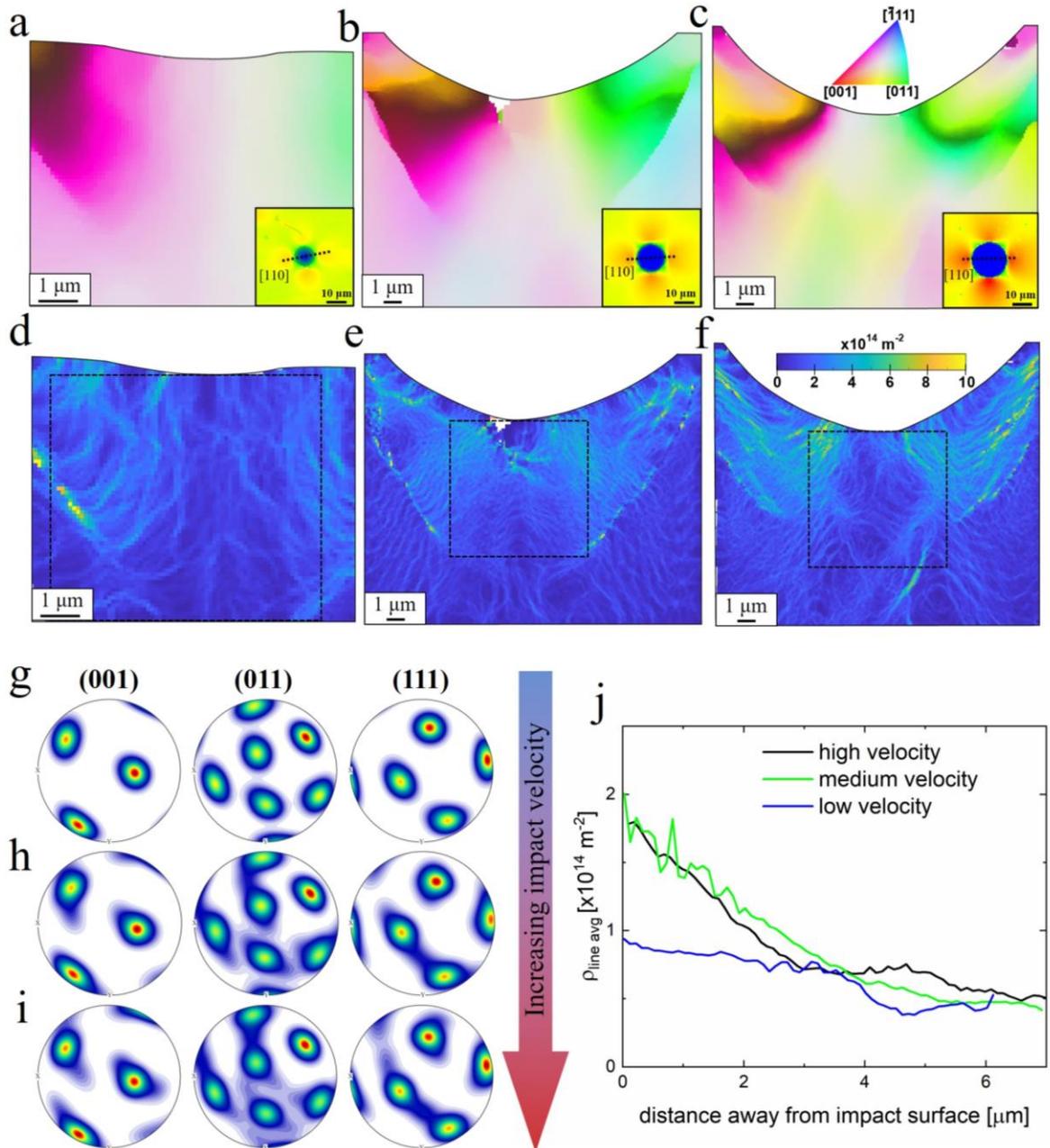

**Figure 5.** Impact-induced nanostructure evolution (a-c) TKD orientation maps taken beneath craters formed from (a) low-velocity (125 m/s), (b) medium-velocity (353 m/s), and high-velocity (550 m/s) impacts, (d-f) dislocation density maps from (d) low, (e) medium, and (f) high velocity impacts. (g-i) pole figures corresponding to (d) low, (e) medium, and (f) high velocity impacts, (j) line average dislocation density profiles as a function of distance away from the impact plane (constant areas of analysis identified in (d-f)).

Interestingly, for all impact velocities, a dislocation density gradient having dislocation density decreasing with increasing depth is observed, until ~6 μm from the impact surface, where the dislocation density plateaus at ~5×10$^{13}$ m$^{-2}$ (Fig.5(j)). The medium and high velocity cases follow a very similar trend. In contrast to the low velocity impact with a relative gentle decrease, a sharp decrease of dislocation density away from the impact surface are observed for both medium and high velocity impacts. At 3-4 μm away from impact surface, the decrease of dislocation density with depth becomes much slower and the average dislocation density approaches to the plateau value similar to the low velocity impact case.

### 3.4 Dynamic hardness evolution under impact

Microprojectile impact testing has been demonstrated to be an effective approach to measure dynamic hardness of materials at ultra-high strain rates which is challenging with traditional testing approaches [15]. The dynamic hardness ($H_D$) of a material is calculated as the ratio of plastic work done by projectile impact ($W_{plastic}$) to the impact-induced plastic deformation volume ($V_{impact}$, volume of crater below the substrate baseline) as follows,

$$H_D = \frac{W_{plastic}}{V_{impact}} = \frac{\frac{1}{2}m_p(V_i^2 - V_r^2)}{V_{impact}}$$

where the plastic work is equal to the kinetic energy loss of projectile during impact. The effects of crystal orientations on the measured hardness of materials under high-speed projectile impacts at different velocities are shown in Fig.6(a). With increasing impact velocity from 120 m/s to ~360 m/s, the dynamic hardness increases for different crystal orientations following slightly different scaling, which is consistent with the previous results on strain rate dependent material hardness of single-crystal Al [55] and polycrystalline metals, *e.g.* iron and copper [15]. During impact, the increased contact force in substrate as well as the possible nucleation mechanism transition, lead to a higher density of dislocation nucleation, being confirmed by the TKD measurement in previous section. This increased dislocation density results in higher material hardness [55]. The measured dynamic hardness of (111) substrates are slightly higher than (011) and (001) substrates, which is consistent with the results obtained from quasistatic indentation of Al single-crystal substrates in our present study as well as reported in prior studies [35]. The indentation hardness of material is directly related to the resolved shear stress on the active slip systems as described by Schmid's law [26,56,57]. For impact on (111) substrates that have lower maximum Schmid's factor (0.27;

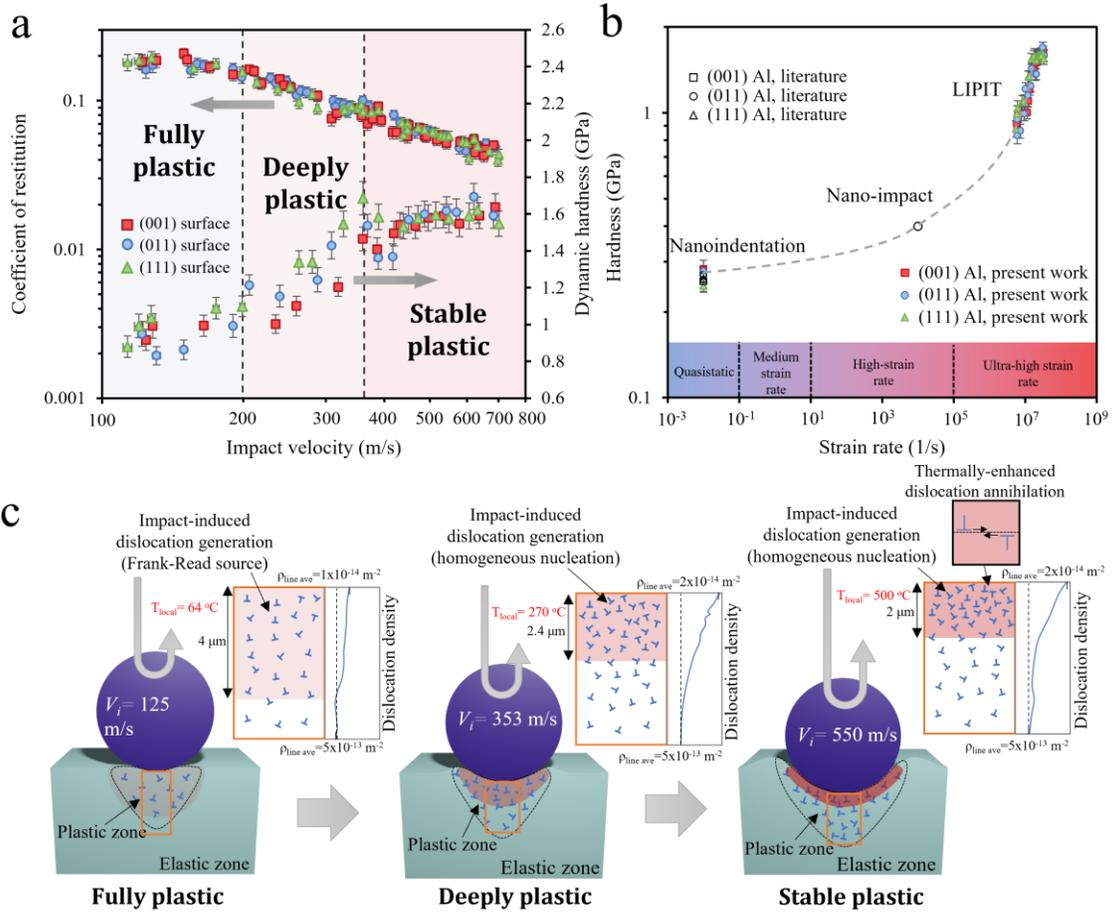

**Figure 6.** (a) Coefficient of restitution and dynamic hardness of single-crystal Al with different crystal orientations as functions of projectile impact velocity. Three regions indicated by different colors represent different deformation mechanisms. (b) Dependency of material hardness on the testing strain rates, from quasistatic ($10^{-2}$ 1/s) to dynamic ($3.3\times10^7$ 1/s) from different testing approaches. (c) Illustration of the proposed microstructural evolution in single-crystal Al substrate subjected to impact at different velocities.

6 slip systems) compared to (001) substrate (0.41; 8 slip systems) and (011) substrate (0.41; 4 slip systems), the activation of multiple slip systems requires higher stress levels, which lead to its higher hardness. With impact velocity increasing higher than ~360 m/s, the dynamic hardness of all three crystal orientations reaches a plateau regime, which correlates well with the dislocation density saturation from TKD results (Fig.5(j)): the medium velocity ($V_i$ =353 m/s) impacted sample investigated by TKD is right at the transition point, and the dynamic hardness and dislocation density distribution are very similar to that of the high velocity case (550 m/s). It may originate from the interplay between the higher density of dislocation from increased deformation strain rate and higher dislocation migration and annihilation from adiabatic heating-induced local temperature rise (up to 670 °C, see SI). Together with the *COR*, we observe three distinct plastic deformation regimes with different deformation mechanisms, in contrast to the previously identified two plastic deformation regimes (fully plastic and deeply plastic). Beyond the fully plastic regime, the deeply plastic regime exhibits increasing dynamic hardness and dislocation density gradient ($V_i$: 200 m/s to ~360 m/s) and the stable plastic regime we identify exhibits relatively constant dynamic hardness and dislocation density gradient ($V_i$: ~360 m/s to 700 m/s).

The effect of strain rate (from quasistatic, $10^{-2}$ 1/s to highly dynamic, $10^7$ 1/s) on hardness of single-crystal Al is shown in Fig.6(b), which includes hardness value obtained from quasistatic indentation (this work and literature [35], high-strain rate indentation (initial strain rate $>10^4$ 1/s) from literature [55], and ultra-high-strain-rate impact from microprojectile impact tests (this work). The strain rate sensitivity of hardness is evident where the hardness at ultra-high strain rate regime is much higher than the quasistatic and low strain rate dynamic regimes. Due to lack of experiments at such intermediate strain rate regime for single-crystal Al, an accurate transition strain rate could not be determined. A similar transition to higher strain rate sensitivity has previously been reported in polycrystalline iron and copper to be ~$10^3$–$10^4$ 1/s [15]. This transition stems from the changes in plastic deformation mechanisms from thermally activated dislocation motion to dislocation drag [7].

Based on our studies, the impact-velocity-dependent deformation mechanisms and microstructural evolutions, especially the dislocation density distribution, in single-crystal Al are shown in Fig.6(c). At low impact velocity ($V_i$ <200 m/s), dislocation generates near to the substrate surface due to projectile impact, from the Frank-Read source nucleation. With increasing impact velocity, the higher strain rate leads to the gradual increase of dislocation density, the formation of a relatively small dislocation density gradient along the depth direction in substrate, and a gentle increase of dynamic hardness with increasing impact velocity. This region is also characterized by the "fully plastic" rebound scaling law of $COR \propto v_i^{-1/2}$, which is related to the rebound initiation position in the projectile/substrate interface. With impact velocity increasing greater than 200 m/s, the higher contact force results in transition of dislocation generation mechanism to homogeneous source nucleation, thereby significantly increasing the near-surface dislocation density which eventually results in a large dislocation density gradient. The rebound initiation position moves deeper into the inner substrate, and follows the "deeply plastic" regime rebound scaling law of $COR \propto v_i^{-1}$. Within this regime (deeply plastic), increasing the impact velocity increases the dislocation density within the material, and leads to the dramatic increase in the dynamic hardness. Further increase in impact velocity beyond 360 m/s, transition the behavior into the "stable plastic" regime. The impact-induced local heating and temperature rise in substrate starts to play a more important role in determining the dislocation density because of thermally-enhanced dislocation motion and annihilation. In this regime, the dislocation density increases from increasing strain rate is fully offset by the enhanced dislocation annihilation from thermal effects, especially in the thermally-affected zone ($T_{local}$ = 500 °C at 550 m/s impact and maximum $T_{local}$ = 670 °C at 700 m/s impact). This leads to the relatively unchanged dislocation density gradient and plateau of dynamic hardness with further increase in impact velocity.

## 4. Conclusions

In summary, we investigated the dynamic behavior of single-crystal Al under ultra-high-strain-rate deformation and showed that the crystal orientation dependent characteristic surface pile-up patterns observed in quasistatic indentation also followed in high-velocity spherical projectile impacts. The observed surface topography correlates well with the characteristic {111} <110> slips that ensue in FCC single crystals, demonstrating primarily a slip dominant plastic deformation mechanism from quasistatic to ultra-high strain rates (up to $10^7$ 1/s). The dynamic hardness exhibits strong dependency on strain rate and crystallographic orientation. The post-mortem TKD analysis on the highly deformed region of the substrate shows impact velocity dependent microstructural evolution, including crystal orientation changes and the formation of a dislocation density gradient. Together with *COR* analysis, we identified a distinct stable plastic regime beyond the previously known fully plastic and deeply plastic deformation regimes with different deformation mechanisms and resultant microstructures. Our study provides new insights on the dynamic hardness evolution stemming from distinct microstructure evolutions of single-crystal FCC metals under high velocity impacts. These findings can enable the development of high-strain rate plasticity models of metals and single-crystal superalloys. We also show that the high-velocity projectile impact onto metals is an effective approach to introduce strong spatial gradient in dislocation density, which can enhance the surface mechanical properties, as it can be employed in shot peening and surface mechanical attrition treatment.

**Acknowledgements**

We acknowledge the partial support for this research provided by the University of Wisconsin-Madison, Office of the Vice Chancellor for Research and Graduate Education with funding from the Wisconsin Alumni Research Foundation. We also acknowledge the partial support from the U. S. Office of Naval Research under PANTHER award number N000142112916 through Dr. Timothy Bentley. The facilities and instrumentation supported by NSF through the University of Wisconsin Materials Research Science and Engineering Center (DMR-1720415) are acknowledged. We also acknowledge Shreya Sreedhar for her support in deformation volume analyses.

**Competing interests**

The authors declare no competing interests.**References**